# Checking and Automating Confidentiality Theory in Isabelle/UTP

Lex Bailey (they/them)[0000-0002-0526-6508] Jim Woodcock[0000-0001-7955-2702]
Simon Foster[0000-0002-9889-9514] Roberto Metere[0000-0001-6992-4285]

University of York, Department of Computer Science
{lex.bailey/jim.woodcock/simon.foster/roberto.metere}@york.ac.uk

**Abstract.** The severity of recent vulnerabilities discovered on modern CPUs, e.g., Spectre [1], highlights how information leakage can have devastating effects to the security of computer systems. At the same time, it suggests that confidentiality should be promoted as a normal part of program verification, to discover and mitigate such vulnerabilities early in development. The theory we propose is primarily based on Bank's theory [2], a framework for reasoning about confidentiality properties formalised in the Unifying Theories of Programming (UTP) [3]. We mechanised our encoding in the current implementation of UTP in the Isabelle theorem prover, Isabelle/UTP [4]. We have identified some theoretical issues in Bank's original framework. Finally, we demonstrate how our mechanisation can be used to formally verify of some of the examples from Bank's work.

## 1 Introduction and Background

Confidentiality is the desirable property of a computer system that ensures secret information remains unknown to users without the authority to know it. It is an essential property for many systems, especially those that hold or process personal information such as passwords, addresses, medical records, and other sensitive information.

Confidentiality should be of equal concern to programmers as correct program behaviour and safety. Various recent examples of confidentiality leaks in modern CPUs suggest that, at least in CPU design, this has not been the case (Metldown [5], Spectre [1], Zenbleed [6]).

Verifying that a program correctly implements a specification is to verify that the program *refines* the specification. Traditional refinement preserves behaviour, but not confidentiality. This is the refinement paradox highlighted by Jeremy Jacob [7]. A more restrictive refinement is required.

This problem has been approached in various ways. Jan Jürjens proposed a mathematical framework that includes security preserving stepwise refinement operations [8]. Carroll Morgan describes an approach that formalises *ignorance* and preserves it during refinement [9]. More recently, Michael Banks proposed a detailed framework for specifying and checking confidentiality properties and preserving them during refinement [2].

Our contribution is a formal verification tool implemented in the Isabelle theorem prover [10] and based on Banks's confidentiality framework [2]. Tool support for Banks's framework has thus far been very limited. Our tool automates verifying confidentiality properties of programs. We have used it to check various examples from Banks's work and our own example case.

Banks's work makes use of Hoare and He's Unifying Theories of Programming (UTP) semantic framework [3]. This makes it appealing to us, as it means that Banks's framework should work well with tools that already support UTP notations and semantics.

We based our tool on Isabelle/UTP [4], which is a distribution of Isabelle along with a layer of support for UTP. We benefit from the various automated proof tools that Isabelle offers and show that they can complete proofs on examples from Banks's work.

Completing this work has, by necessity, included checking that Banks's theory is sound. As detailed in the next section, we found errors in Banks's work that require us to rework parts of the theory.

## 2  An Example System

We present a small example specification for a system that handles user authentication by username and password. This paper describes the example informally. The full system specification, analysis and proofs are available on github.[1]

The system has a database of usernames and passwords. The database is a partial function from usernames to passwords. Authenticating a user involves determining if the provided username and password are in the database. First by looking up the password given the username (if it exists) and then by comparing the provided password to the stored password.[2]

Regardless of success or failure, the system should provide a message to the user to provide them feedback on the result of the authentication process. This specification does not require any specific text in each success message or error message.

---

[1] https://github.com/lexbailey/Banks/blob/published_state/examples/user_pass.thy

[2] This is not best practice. We are presenting a simplified system to ease understanding. Best practice is to only store password hashes, not passwords.

Suppose the confidentiality property that interests us is that the list of usernames remains totally unknown to any user that is not in the database. Clearly a user in the database knows that they exist in the database, but an attacker without an username in the database should not know of any entries, nor have a way to determine any of them.

Let us now consider two different implementations of this specification. The first one provides three different messages for the three different possible executions of the authentication program.

> User exists, password correct → "Success"
> User exists, password incorrect → "Incorrect password"
> User does not exist → "No such user"

And the second system only has two possible messages for these three cases:

> User exists, password correct → "Success"
> User exists, password incorrect → "Authentication failed"
> User does not exist → "Authentication failed"

Both implementations refine the specification, but only the second one preserves the confidentiality property. For the first system, the attacker can determine if "Alice" is a user in the database simply by trying to authenticate as Alice with any password. If the message they receive is "No such user", then they know that Alice is not a user in the database, otherwise Alice must be in the database.

## 3   Views, Localisation, Globalisation, and Inference

The first concept introduced in Banks's work is the concept of a view: a user's interface to a system. In this context, the system is a UTP program. As such, it has an alphabet of system variables and is described by a relation over members of the alphabet. Each pair of input and output states in the relation is a valid execution of the program. Output variables in the system's alphabet are denoted with a dash. (input variable $x$ corresponds to the output variable $x'$)

In Banks's work: views are written just like UTP programs. They have alphabets, with variables that can represent exactly what information the user can observe, but their domain is the union of this alphabet and the system alphabet. In this way, a view describes a relation between system states and view states. Below are some example views from page 24 of Banks's thesis [2].

$$V1 \stackrel{\text{def}}{=} a_1 = x \wedge b_1 = y$$
$$V2 \stackrel{\text{def}}{=} c_2 = \max(x, y)$$
$$V3 \stackrel{\text{def}}{=} d = x \wedge (e_3 = 0 \triangleleft x < y \triangleright e_3 = 1)$$

In these examples, $x$ and $y$ are system variables. All the other variables are view variables. $V1$ allows the user to inspect the variables $x$ and $y$ directly. $V2$ allows the user to know the maximum of $x$ and $y$, and $V3$ allows them to see the exact value of $x$, and a flag ($e_3$) set when $x \geq y$.

Banks defines localisation ($\boldsymbol{L}$), globalisation ($\boldsymbol{G}$), and inference ($infer$) functions for transforming views. Localisation of a system, with respect to a view, hides the system variables and provides a predicate that encodes all possible observations of the system through the view.

$$\boldsymbol{L}(V, P) \stackrel{\text{def}}{=} \exists x, x' \cdot \Delta V \wedge P$$

Globalisation is the compliment of localisation, and reconstructs the system behaviour that can be determined only from the localised view. It is not an exact inverse of localisation because localisation is lossy. Banks explains this more formally: "The relation $\boldsymbol{G}(V, U)$ is the weakest specification of a system such that each behaviour of $\boldsymbol{G}(V, U)$ projects through $V$ to an interaction of $U$."

$$\boldsymbol{G}(V, U) \stackrel{\text{def}}{=} \forall x_V, x'_V \cdot \Delta V \Rightarrow U$$

Finally, the inference function takes a system behaviour, a view, and an observation made through that view, and encodes what a user can learn about the system behaviour given that observation. It is the weakest precondition that establishes the observation $\psi$ of the system $P$ through the view $V$.

$$infer(P, V, \psi) \stackrel{\text{def}}{=} P \wedge \neg \boldsymbol{G}(V, \neg \psi)$$

## 4 Encoding and Checking Banks Confidentiality Theory

This section details our encoding of Banks's theory in Isabelle/UTP. It includes details of the Isabelle code, and it explains errors found in the original work and corrections or changes required to accommodate those.

### 4.1 Views

As we are using Isabelle/UTP to encode and check this work, we can write views in a very similar style to Banks's notation. The key difference is that we use a hierarchy of alphabets to separate the view and system variables.

Using a hierarchy in this way enables us to refer to the entire set of view variables as if they were a single variable and the entire set of system variables as if they were another single variable, which we will need later.

In Isabelle/UTP syntax:

```
alphabet ('s, 'v) viewed_system =
  sys :: 's
  vu :: 'v
```

This defines a type called `viewed_system` which is a record with two fields. The first is `sys`, which represents the system variables, and the second is `vu` which represents the view variables. These fields can have any types, but typically they will also be record types created with the **alphabet** command.

With this we can define views in a similar way to Banks's examples. Below is a recreation of the $V1$ example view from earlier, along with the required alphabet definitions.

```
alphabet sys_vars =
  x :: int
  y :: int

alphabet v1_view_vars =
  a1 :: int
  b1 :: int

definition V1 where "V1 = (
    vu:a1 = sys:x ∧ vu:b1 = sys:y
  )ₑ"
```

$V1$ has the type `(sys_vars, v1_view_vars) viewed_system ⇒ Bool`, which can be seen as a predicate for valid observations or a relation between system states and view states. Notice the use of `()ₑ` syntax, which is part of Isabelle/UTP, specifically the shallow-expressions module [11]. An expression inside e-brackets is transformed into a predicate.

Banks's original work uses a flat namespace for all variables, as is normally done in UTP. However, since the system and view variables exist in their respective namespaces in our encoding of Banks's work, we have enabled the use of the same name between system and view. This could be an advantage, as in the case of $V1$, we could rename `vu:a1` to `vu:x` and `vu:b1` to `vu:y` to match the system variables that they correspond to without introducing any ambiguity since they always represent the same value.

Banks defines several healthiness conditions for views. The two most important ones, for now, are **VH1** and **VH2**. Various later theorems only hold for views for which these healthiness conditions hold. Banks defines **VH1** as follows: "A view $V$ is **VH1**-healthy if and only if $V$ maps each behaviour to *at least one* interaction". Where "behaviour" and "interaction" refer to system states and view states, respectively. The formal definition is as follows.

$$VH1(V) \stackrel{\text{def}}{=} (\exists x_V, x'_V \cdot V) \Rightarrow V$$

Here, $x_V$ and $x'_V$ represent all the view variables before and after execution (respectively). A more intuitive way to read this condition is: the view does not restrict the system. (Because if it did, there would be an otherwise valid system state, but no related view state.)

In our encoding, the definition of **VH1** is slightly different to Banks's definition.

```
definition VH1
  where "VH1 V = ((∃ vu • V) ⟶ V)ₑ"
```

This definition appears to drop the requirement for some assignment of the "after" variables (the dashed variables) to exist. We mainly treat views as if they only have one copy of their variables rather than an undashed copy and a dashed copy. This simplifies much of the encoding without loss of generality. Our definition does not mention dashed variables because they do not exist in our view definitions.

Banks also defines another healthiness condition **VH2**. In Banks's words: "A view $V$ is **VH2**-healthy if and only if its alphabet contains only undashed variables"

$$VH2(V) \stackrel{\text{def}}{=} \exists x', x'_V \cdot V$$

Our views only have undashed variables by definition, and thus, we do not need **VH2**. In our Isabelle code, we enforce this through the type system. It is impossible to write a view that would not comply with **VH2**. For completeness, however, our encoding includes a trivial definition for **VH2**:

```
definition VH2 where "VH2 = id"
```

To finish justifying why we can ignore dashed variables, let's look at the next function that Banks introduces. This is the $\Delta$ function.

$$\Delta V \stackrel{\text{def}}{=} V \wedge V'$$

$V'$ represents $V$ with all undashed variables renamed as dashed variables. Many of the functions defined later in Banks's work use $\Delta V$ rather than $V$ when operating on a view.

The purpose of **VH2** and $\Delta$ is to make sure that the user's view of the system is consistent between the before and after states. The view that the user has before the program starts is the same one they will have after the program has ended.

We encode **VH3** as per Banks's definition. This is the "functional view" healthiness condition. This is not required for every view but simplifies some proofs when it holds.

$$\mathbf{VH3}(V) \stackrel{\text{def}}{=} (\exists_1 x_V, x'_V \cdot V) \Rightarrow V$$

```
definition VH3
  where "VH3 P = (P !\\ $vu → P)ₑ"
```

**VH** is a healthiness condition that combines both **VH1** and **VH2**. Since **VH2** reduces to $id$ in our encoding, we show **VH** to be the same as just **VH1**.

We also encoded and checked various theorems that Banks provided around views and their healthiness conditions.[3]

### 4.2   Calculating Interactions

We encoded the localisation, globalisation, and inference functions from Banks's work, as part of our tool. As a reminder, localisation is defined as follows:

$$L(V, P) \stackrel{\text{def}}{=} \exists x, x' \cdot \Delta V \wedge P$$

Our encoding of this function closely resembles the original, with two notable differences.

```
definition L
  where "L V P = (
    ∃ (sys˂, sys˃) • (Δ V ∧ P ↑ sys²)
  )ₑ"
```

The first is because Isabelle/UTP uses superscript "<" and ">" symbols to denote the *before* and *after* variables, respectively. The second difference is the additional alphabet extrusion $\uparrow$ `sys`², which takes a relation over some type `'s` and turns it into a relation over the type `('s, 'v) viewed_system`. This is to make sure that it unifies with the type of the view after applying Δ.

---

[3] Specifically: We checked Laws 3.8, 3.9, 3.10, Corollary 3.12, and Lemma 3.13. Laws 3.9, 3.10, and 3.12 trivially hold because of our simplification of **VH2**.

This up-arrow notation for alphabet extrusion is used in various places in our work, along with the opposing down-arrow notation for alphabet restriction, which extracts smaller types from larger ones.

Our Isabelle/UTP encoding of the globalisation function also closely resembles Banks's definition.

$$G(V, U) \stackrel{\text{def}}{=} \forall x_V, x'_V \cdot \Delta V \Rightarrow U$$

```
definition G where "G v u = (∀ (vu‹, vu›) • ((Δ v) ⟶ u))ₑ"
```

Finally, the `infer` function is also encoded.

$$infer(P, V, \psi) \stackrel{\text{def}}{=} P \wedge \neg G(V, \neg \psi)$$

```
definition infer where "infer P V ψ = (P ∧ ¬ G V (¬ ψ))ₑ"
```

With these definitions, we have enough to reproduce all the example uses of these functions from Banks's work.

### 4.3 Localisation and Globalisation Examples

We have reproduced Banks's examples of using Localisation to check all the ways that a user can interact with a system[4].

Banks provided an example system:

$$Ex \stackrel{\text{def}}{=} x \geq 0 \wedge y \geq 0 \wedge x + y = 10$$

Our encoding of Banks's framework enables us to write this example system in a similar manner.

```
alphabet sys_vars =
  x :: int
  y :: int

definition ExSys where
  "ExSys = Δ (x ≥ 0 ∧ y ≥ 0 ∧ x + y = 10)ₑ"
```

The theoretical world allowed Banks to be very flexible with types. Isabelle imposes strict typing. This is why we had to use Δ to turn the example system

---

[4] These are from page 24 of Banks's thesis

from a condition into a relation. Our implementation of the **L** function only allows relations.

Now we can use the **L** function, and the example views introduced earlier, and check that Banks correctly calculated the space of user interactions with this example system through these views.

Starting with $V1$, Banks calculates the space of interactions as follows:

$$V1 \stackrel{\text{def}}{=} a_1 = x \wedge b_1 = y$$
$$L(V1, Ex) = a_1 \geq 0 \wedge b_1 \geq 0 \wedge a_1 + b_1 = 10$$

This is a fairly simple calculation, as it just maps the variables $x$ and $y$ directly to $a_1$ and $b_1$ respectively. We can check that this holds using the basic automatic proof tool "`pred_auto_banks`" which is a wrapper around Isabelle/UTP's "`pred_auto`" tool but with additional definitions and lemmas for Banks's framework.

```
alphabet v1_view_vars =
  a1 :: int
  b1 :: int

definition V1 where "V1 = (
    vu:a1 = sys:x ∧ vu:b1 = sys:y
  )ₑ"

lemma v1_healthy1: "V1 is VH"
  by (pred_auto_banks add: V1_def)

definition Loc1 where "Loc1 = L V1 ExSys"

lemma localise_v1 :
  "Loc1 = Δ((a1 ≥ 0 ∧ b1 ≥ 0 ∧ a1 + b1 = 10)ₑ ↑ vu)"
  by (pred_auto_banks add: Loc1_def V1_def ExSys_def)
```

This example is the simplest of the three. We introduce the alphabet first, then define the view. Lemma `v1_healthy1` shows that `V1` is a healthy view. `Loc1` is the projection of `ExSys` through `V1` as defined by `L`, and finally the lemma `localise_v1` shows that `Loc1` is the same as Banks's calculation.

For the next two examples, we'll skip directly to the proof that Banks's calculation is correct:

$$V2 \stackrel{\text{def}}{=} c_2 = \max(x, y)$$
$$L(V2, Ex) = c_2 \geq 5 \wedge c_2 \leq 10$$

```
lemma localise_v2 : "Loc2 = Δ((c2 ≥ 5 ∧ c2 ≤ 10)ₑ ↑ vu)"
  apply (pred_auto_banks add: Loc2_def V2_def ExSys_def)
  by presburger+
```

$$V3 \stackrel{\text{def}}{=} d = x \wedge (e_3 = 0 \triangleleft x < y \triangleright e_3 = 1)$$
$$\boldsymbol{L}(V3, Ex) = (d_3 \geq 0 \wedge d_3 < 5 \wedge e_3 = 0) \vee (d_3 \geq 5 \wedge d_3 \leq 10 \wedge e_3 = 1)$$

```
lemma localise_v3 : "Loc3 = Δ((
    (d3 ≥ 0 ∧ d3 < 5 ∧ e3 = 0)
   ∨(d3 ≥ 5 ∧ d3 ≤ 10 ∧ e3 = 1)
   )ₑ ↑ vu)"
  apply (pred_auto_banks add: Loc3_def V3_def ExSys_def)
  by presburger+
```

For these two proofs Isabelle's Sledgehammer[5] was able to quickly find that using the presburger tactic completed the proof. This level of automation makes our tool very convenient for the end user.

Finally, we show that the example applications of the globalisation function $\boldsymbol{G}$ are also correctly calculated by Banks. Indeed, it is the case that these are all correct.

$$\boldsymbol{G}(V1, \boldsymbol{L}(V1, Ex)) = x \geq 0 \wedge y \geq 0 \wedge x + y = 10$$

```
lemma "(G V1 Loc1) = Δ((x ≥ 0 ∧ y ≥ 0 ∧ x + y = 10)ₑ ↑ sys)"
  by (pred_auto_banks add: Loc1_def V1_def ExSys_def)
```

$$\boldsymbol{G}(V2, \boldsymbol{L}(V2, Ex)) = \max(x, y) \geq 5 \wedge \max(x, y) \leq 10$$

```
lemma "(G V2 Loc2) =
  Δ(((max x y ≥ 5) ∧ (max x y ≤ 10))ₑ  ↑ sys)"
  apply (pred_auto_banks add: Loc2_def V2_def ExSys_def)
  by presburger
```

$$\boldsymbol{G}(V3, \boldsymbol{L}(V3, Ex)) = (x \geq 0 \wedge x < 5) \triangleleft x < y \triangleright (x \geq 5 \wedge x \leq 10)$$

```
lemma "(G V3 Loc3) =
  Δ(( if x < y then (x ≥ 0 ∧ x < 5) else (x ≥ 5 ∧ x ≤ 10))ₑ
  ↑ sys)"
  apply (expr_simp_banks add: Loc3_def V3_def ExSys_def)
  apply (pred_simp)
```

---

[5] Sledgehammer is a feature of Isabelle that wraps the various automatic proof methods in Isabelle for one-click proof automation and reconstruction.

```
by presburger+
```

Once again, the proofs were mostly automatic. The only tricky part is that currently the "if-then-else" operator does not work well with `pred_auto`, and so `expr_simp` is needed first for the case of $V3$. This is a point where automation could be improved in future.

### 4.4 Inference Example

The next example from Banks's work contains a small error. This is the "guessing game" example[6]. In this example, a player (Alice) guesses a number between 1 and 10. The example demonstrates the `infer` function, which can tell us what we learn about the secret number based on the guess and the response to the guess as generated by the logic of the game.

The game and the user's view of the game are defined by Banks as follows:

$$GUESS_0 \stackrel{\text{def}}{=} n \in 1..0 \land g \in 1..10 \land \begin{pmatrix} & g > n \Rightarrow r' > 0 \\ \land & g = n \Rightarrow r' = 0 \\ \land & g < n \Rightarrow r' < 0 \end{pmatrix}$$

$$ALICE \stackrel{\text{def}}{=} g_A = g \land r_A = r$$

The error we found is a minor one. Banks uses this guessing game in an example use of the infer function, as follows:

$$infer(GUESS_0, ALICE, g_A = 7 \land r'_A > 0)$$
$$= n \in 8..10 \land g = 7 \land r' > 0$$

This is not correct. The error can be fixed in two ways. If the example was supposed to show what happens when $r$ is greater than 0, then the value of $n$ must be strictly less than $g$, and so it should be in the range 0..6.

$$n \in 0..6 \land g = 7 \land r' > 0$$

If the example was supposed to show the condition under which you can infer that n must be in the range 8..10, then $r'$ should actually be less than 0

$$n \in 8..10 \land g = 7 \land r' < 0$$

Both of these fixes assume that the initial definition of $GUESS_0$ is as intended. We chose to show that the example is correct with the $r' < 0$ fix.

---

[6] Example 3.30 on page 33 of Banks's thesis [2]

```
lemma
  "(
    (infer (Guess0) (Alice) (
      (ga˙ = 7 ∧ ra` < 0)ₑ ↑ vu²
    ))
    =
    (g˙ = 7 ∧ r` < 0 ∧ n˙ ≥ 8 ∧ n˙ ≤ 10)ₑ  ↑ sys²
  )"
  by (pred_auto_banks add: Alice_def Guess0_def)
```

## 4.5  Designs

After introducing Views, Banks goes on to extend the concept to the UTP theory of Designs. A design is an extension of the concept of a program specification as a predicate. It has a precondition and a postcondition, and two additional special variables: *ok* and *ok′*. *ok* indicates that the program has started, and *ok′* indicates that it has terminated.

Views are denoted with the turnstile operator, and defined for a precondition *pre* and postcondition *post* as follows:

$$pre \vdash post \stackrel{\text{def}}{=} ok \wedge pre \Rightarrow ok' \wedge post$$

Our encoding of this part of Banks's work also has some differences from the original work. We have applied some simplifications similar to those we applied to **VH2**.

To make views for designs work like regular UTP designs, Banks introduces a variable called $ok_V$ and a healthiness condition that ensures this variable matches the *ok* variable of the system's design alphabet.

$$\boldsymbol{OK}(V) \stackrel{\text{def}}{=} V \wedge ok_V = ok$$

This means that there is a second copy of the *ok* variable which must always match the original for healthy designs. In our encoding it is more convenient to have only one copy of the *ok* variable. For this reason, we need not introduce $ok_V$ at all. This removes the need for the **OK** healthiness condition.

To justify this, we can examine the hierarchy of alphabets used in a viewed design. At the top level is the `des_vars` alphabet, which contains only the *ok* variable. This is then extended with both the view alphabet and the system alphabet in the form of the parameterised `viewed_system` type. An example is shown in Figure 4.1.

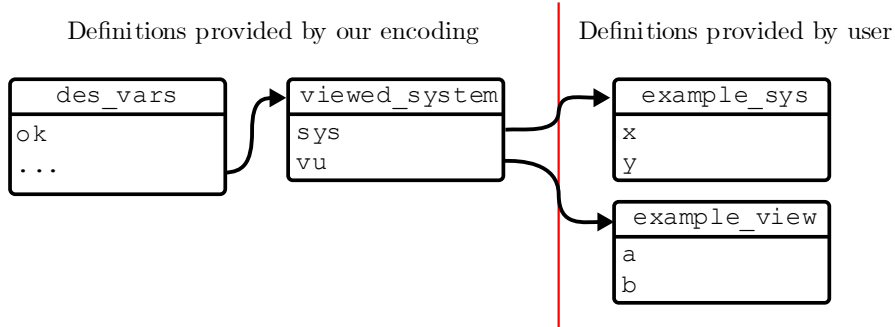

*Figure 4.1 An example of the alphabet type hierarchy for a viewed design with system variables x,y and view variables a,b.*

The type for the alphabet of a view is of the form `('s, 'v) viewed_system des_vars`. We refer to a relation between pairs of this type as a viewed design.

Since the *ok* variable is not directly associated with the system, the *ok* variable in the design can easily be split out when needed. This separation enables the deduplication of *ok*.

**VHD** is a healthiness condition that combines **VH** and **OK**, and reduces to just **VH** in our encoding. Since **VH** is just **VH1**, we also show that **VHD** reduces to **VH1**.

Banks introduces a definition for a "Local design" as follows:

$$Pre \vdash_V Post \stackrel{\text{def}}{=} ok_V \land Pre \Rightarrow ok'_V \land Post$$

Where *Pre* and *Post* are predicates with alphabet $x_V, x'_V$. This is the view variables alphabet. The system variables are absent. This encodes what a user knows about the output state for a given input state.

Given the structure of the types we use to represent these alphabets, adding in the *ok* variable to a predicate over the view variables does not require a special definition of a design. Since we have done away with $ok_V$ and use *ok* in its place (because they are always equal for healthy viewed designs) we can achieve the same thing with the normal definition of designs provided by Isabelle/UTP.

Banks claims the following to be true for any **VHD** healthy view *V*. (Lemma 3.35)

$$L(V, Pre \vdash Post) = \neg L(V, \neg Pre) \vdash_V L(V, Post)$$

We show that this does not hold in general. Our proof is provided in the appendix. (Sect. 8) While most of the steps of Banks's proof of Lemma 3.35 were correct, a crucial step inadvertently introduced an additional requirement for the view to accept any observation. Such a view is not useful as it can only model a user that cannot determine anything about the system.

This cuts short our endeavour to check and implement Banks's work. The remainder of the work builds on this foundation. Without this part of the framework, we need to create a different method for modelling confidentiality. Our future work will be inspired by the concepts from Banks's work, but will differ greatly in the details.

## 5  Related Work

This is not the first endeavour to mechanise Banks's work to leverage automatic theorem provers for verifying confidentiality. Ibrahim Shiyam mechanised part of Banks's work [12]. Their implementation differs from ours in various ways.

Shiyam's mechanisation uses a custom tool to parse a specification and generate a file to import into Isabelle for checking. Our tool instead makes use of Isabelle/UTP to enable the specification to be written in Isabelle directly, simplifying the toolchain for a more convenient workflow.

There is also a more recent (2021) proposed framework for modelling information flow in UTP. Mu Chunyan and Li Guoqiang introduced a framework [13] that achieves similar goals to Banks's work, but in a different way. Their framework does not have views in the same way as Banks's framework. Instead, the observability of program state is encoded by assigning a security level to each variable in the program, and then hiding variables above a given security level.

While it would be interesting to implement Mu and Li's work in Isabelle/UTP, it is not obvious how one would begin doing this. In particular, it is not easy to describe a system where each variable has a matching security-level variable and still get the advantages that Isabelle/UTP offers.

## 6  Conclusions and Future Work

Our encoding of Banks's confidentiality theory provides the beginnings of a user-friendly system for analysing confidentiality in systems modelled in a UTP style. We have identified some errors in Banks's work, some of which require us to investigate alternative paths to having a complete system.

We will continue to develop the mathematics and tooling to widen the range of confidentiality properties that it can handle, and to improve the level of automation that it offers. We also plan to develop this theory to be able to model probabilistic behaviours of systems. This will enable us to find confidentiality flaws that would be missed by Banks's framework.

## 7 Where to find our code

The code described in this paper is (at the time of publication) available on github. The Isabelle/UTP code for our encoding of Banks's framework, and our usage examples, can be found at https://github.com/lexbailey/Banks/ The version of the encoding as of the time of publication can be found at the git tag "published_state" https://github.com/lexbailey/Banks/tree/published_state

Isabelle/UTP is required to make use of this encoding. It can be found at https://github.com/lexbailey/isabelle_utp_full and more details about it can be found at the project website at https://isabelle-utp.york.ac.uk/

## 8 Appendix: Proof that Lemma 3.35 does not hold

First we transform the left side. This is mostly as Banks did in his attempted proof:

$$\begin{aligned}
\boldsymbol{L}(V, Pre \vdash Post) &= \exists x, x' \cdot \Delta V \land (\text{ok} \land Pre \Rightarrow \text{ok}' \land Post) \\
&= \exists x, x' \cdot \Delta V \land \bigl(\neg ok \lor \neg Pre \lor (ok' \land Post)\bigr) \\
&= \bigl(\exists x, x' \cdot \Delta V \land (\neg ok \lor \neg Pre)\bigr) \lor \bigl(\exists x, x' \cdot \Delta V \land (ok' \land Post)\bigr) \\
&= \bigl(\exists x, x' \cdot \Delta V \land (\neg ok \lor \neg Pre)\bigr) \lor (ok' \land \exists x, x' \cdot \Delta V \land Post)
\end{aligned}$$

This should be equal to the right side, which we will transform similarly to Banks's proof but in reverse:

$$\begin{aligned}
&\neg \boldsymbol{L}(V, \neg Pre) \vdash_V \boldsymbol{L}(V, Post) \\
&= \bigl(ok_V \land \neg \boldsymbol{L}(V, \neg Pre)\bigr) \Rightarrow \bigl(ok'_V \land \boldsymbol{L}(V, Post)\bigr) \\
&= \neg \bigl(ok_V \land \neg \boldsymbol{L}(V, \neg Pre)\bigr) \lor \bigl(ok'_V \land \boldsymbol{L}(V, Post)\bigr) \\
&= \bigl(\neg ok_V \lor \boldsymbol{L}(V, \neg Pre)\bigr) \lor \bigl(ok'_V \land \boldsymbol{L}(V, Post)\bigr) \\
&= (\neg ok_V \lor \exists x, x' \cdot \Delta V \land \neg Pre) \lor (ok'_V \land \exists x, x' \cdot \Delta V \land Post)
\end{aligned}$$

Having transformed these expressions, we expect them to be equal if Banks's claim is correct.

$$\begin{aligned}
&\bigl(\exists x, x' \cdot \Delta V \land (\neg ok \lor \neg Pre)\bigr) \lor (ok' \land \exists x, x' \cdot \Delta V \land Post) \\
&= (\neg ok_V \lor \exists x, x' \cdot \Delta V \land \neg Pre) \lor (ok'_V \land \exists x, x' \cdot \Delta V \land Post)
\end{aligned}$$

We are only interested in cases where $ok = ok_V$, so we replace all $ok_V$ with $ok$ and simplify the problem.

$$\begin{aligned}
&\bigl(\exists x, x' \cdot \Delta V \land (\neg ok \lor \neg Pre)\bigr) \lor (ok' \land \exists x, x' \cdot \Delta V \land Post) \\
&= (\neg ok \lor \exists x, x' \cdot \Delta V \land \neg Pre) \lor (ok' \land \exists x, x' \cdot \Delta V \land Post)
\end{aligned}$$

Thus the claim is only true if:

$$(\exists x, x' \cdot \Delta V \wedge (\neg ok \vee \neg Pre)) = (\neg ok \vee \exists x, x' \cdot \Delta V \wedge \neg Pre)$$

The variable $ok$ must be either $true$ or $false$. When it is $true$, the claim holds. When it is $false$, the claim reduces as follows:

$$(\exists x, x' \cdot \Delta V \wedge (\neg ok \vee \neg Pre)) = (\neg ok \vee \exists x, x' \cdot \Delta V \wedge \neg Pre)$$
$$(\exists x, x' \cdot \Delta V \wedge (\neg false \vee \neg Pre)) = (\neg false \vee \exists x, x' \cdot \Delta V \wedge \neg Pre)$$
$$(\exists x, x' \cdot \Delta V \wedge (true \vee \neg Pre)) = (true \vee \exists x, x' \cdot \Delta V \wedge \neg Pre)$$
$$(\exists x, x' \cdot \Delta V) = true$$

So the claim only holds if $\exists x, x' \cdot \Delta V$ is universally true. This is not an assumption of the claim, nor would it be valid to add it as an assumption of the claim. Given that a view is only healthy when it does not mention the system variables $x$ and $x'$, this claim only holds when $V = true$. This view models a user who can see nothing, and is the only view for which the lemma holds.